\documentclass[a4paper]{article}
\usepackage[english]{babel}
\usepackage[utf8x]{inputenc}
\usepackage[T1]{fontenc}

\usepackage[a4paper,top=3cm,bottom=2cm,left=3cm,right=3cm,marginparwidth=1.75cm]{geometry}

\usepackage{amsmath}
\usepackage{mathtools}
\usepackage{float}
\usepackage{amsfonts}
\usepackage{graphicx}
\usepackage[colorinlistoftodos]{todonotes}
\usepackage[colorlinks=true, allcolors=black]{hyperref}
\usepackage{cite}
\usepackage{verbatim}

\usepackage{algorithm}
\usepackage[noend]{algpseudocode}
\usepackage{authblk}


\title{\textbf{A Time-of-Flight Imaging System Based on Resonant Photoelastic Modulation}}

\author[1*]{Okan Atalar}
\author[2]{Rapha\"{e}l Van Laer}
\author[2] {Christopher J. Sarabalis} 
\author[2] {Amir~H.~Safavi-Naeini}
\author[1] {Amin Arbabian}

\affil[1]{Department of Electrical Engineering, Stanford University, Stanford, California 94305, USA}
\affil[2]{Department of Applied Physics and Ginzton Laboratory, Stanford University, Stanford, California 94305, USA}
\affil[*]{Corresponding author: okan@stanford.edu}

\begin{document}
\maketitle

\setcounter{secnumdepth}{0}
\section{Abstract}
A time-of-flight (ToF) imaging system is proposed and its working principle demonstrated. To realize this system, a new device, a free-space optical mixer, is designed and fabricated. A scene is illuminated (flashed) with a megahertz level amplitude modulated light source and the reflected light from the scene is collected by a receiver. The receiver consists of the free-space optical mixer, comprising a photoelastic modulator sandwiched between polarizers, placed in front of a standard CMOS image sensor. This free-space optical mixer downconverts the megahertz level amplitude modulation frequencies into the temporal bandwidth of the image sensor. A full scale extension of the demonstrated system will be able to measure phases and Doppler shifts for the beat tones and use signal processing techniques to estimate the distance and velocity of each point in the illuminated scene with high accuracy. 

\setcounter{secnumdepth}{5}

\section{Introduction}
The human visual system and standard image sensors form high-resolution images of their surroundings. These systems are effective in forming images of the surrounding scene but do not provide accurate estimates of depth. Many applications, however, rely on accurate depth images in a scene, including machine vision~\cite{machine_vision_1,machine_vision_2}, tracking~\cite{tracking_1,tracking_2}, autonomous vehicles~\cite{autonomous_vehicles_1,autonomous_vehicles_2,autonomous_vehicles_3} and robotics~\cite{robotics}. The need for generating accurate depth images in a scene necessitates new generation of image sensors. 

Depth imaging in a scene can be achieved through the ToF imaging technique. A scene is illuminated with a controlled light source and the interaction of this light with the scene is captured and processed for estimating the depth in the scene. The most basic method for ToF imaging involves sending a focused beam of light pulse to a particular location in a scene and measuring the time delay of the returned pulse to the optical detector. Scanning the beam allows depth images to be generated. Scanning of the beam can be realized through mechanical~\cite{mechanical_scanning_1,mechanical_scanning_2} scanning  or non-mechanical scanning (solid state). Non-mechanical scanning usually uses optical phased arrays with full control of the phase and frequency of a laser beam~\cite{phased_array_1,phased_array_2,phased_array_3,phased_array_4}, although recently solid state optomechanical steering has also been proposed~\cite{optomechanical}. An alternative method, usually referred to as flash lidar, captures depth images through illuminating a part of the scene with a modulated light source. Flash lidar avoids scanning the beam by capturing a part of the scene at a single shot, making it a possible low cost, fast and effective way of measuring depth images.

One class of flash lidars operate in time domain by measuring the ToF for each sensor pixel after flashing the scene with a light source. Each point in the scene is focused to a specific image sensor pixel with the use of an optical lens. The ToF for the light to arrive at each sensor pixel is used to determine the distance of each point to the sensor. These flash lidars have high unambiguous range and depth resolution, but are limited by cost or spatial resolution since they require pulsed lasers and specialized pixels with high bandwidths~\cite{laser_pulse_camera_1,laser_pulse_camera_2,laser_pulse_camera_3,laser_pulse_camera_4}. Compressed sensing techniques with a single pixel camera and a pulsed laser has also been demonstrated~\cite{compressed_pixel}, but these systems also have limited spatial resolution compared to standard image sensors. 

Another class of flash lidar sends amplitude modulated light to a scene and measures the phase of the reflected light from the scene with respect to the illumination light phase, similar to the operation of stepped frequency continuous wave (SFCW) radar~\cite{SFCW}. This technique has also been referred to as radio frequency interferometry (RFI)~\cite{rfi}, since light is modulated at typical radar operating frequencies and the envelope of the light is used for estimating distances. To detect distances on the order of meters with sub-meter level depth resolution, megahertz modulation frequencies are used. Standard image sensors do not have the bandwidth to capture the phase of megahertz frequencies. The standard method is to demodulate the incoming megahertz frequency to a lower frequency before sampling, similar to the working principle of a superheterodyne receiver. 

\begin{figure*}[t!]
\begin{center}
\includegraphics[width=1\textwidth]{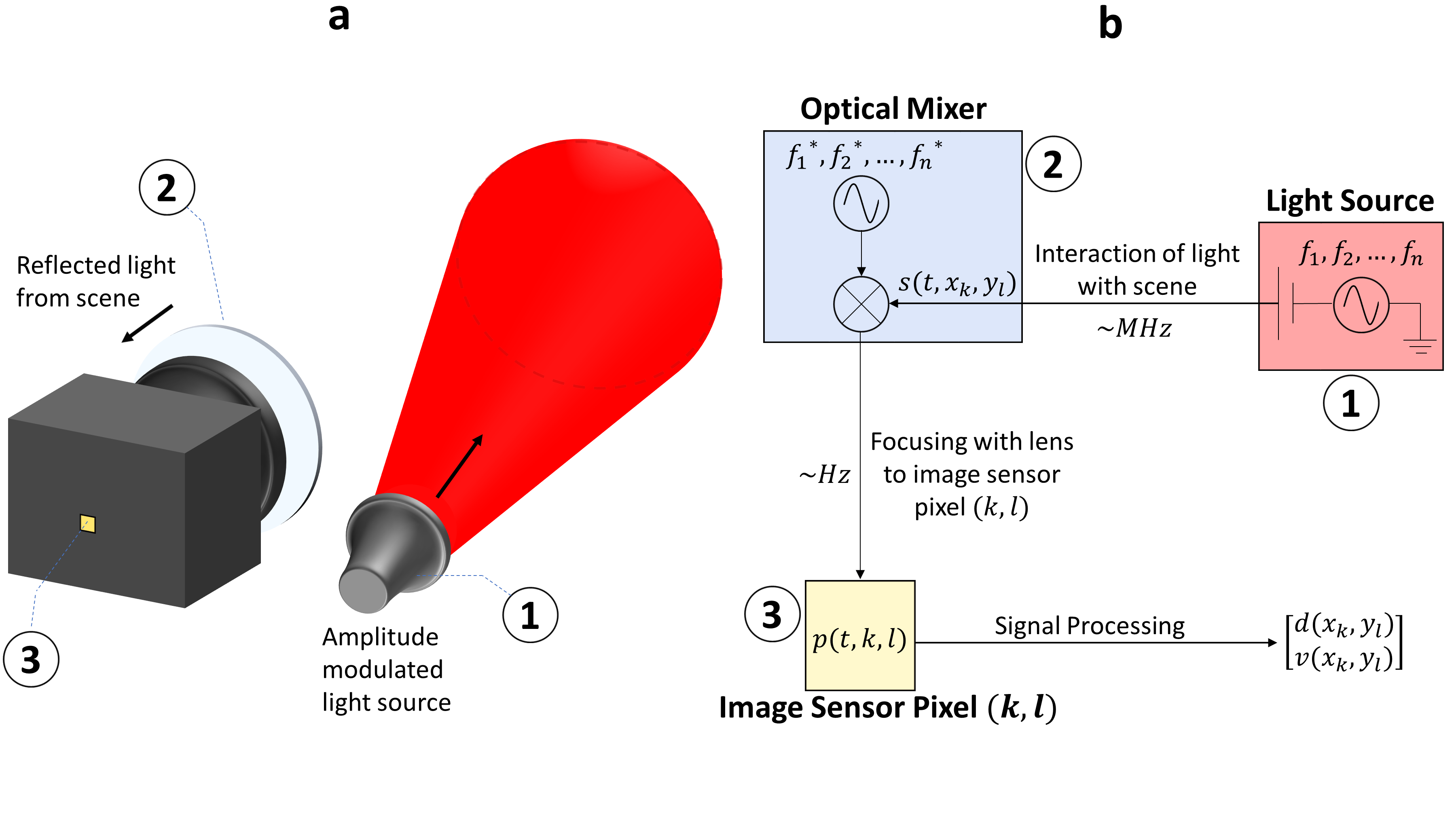}
\caption{\textbf{Flash lidar system to capture distance and velocity in the scene per image sensor pixel (k,l)}. \textbf{a} Schematic of flash lidar system: A light source is amplitude modulated at megahertz frequencies $f_1,f_2,...,f_n$ and a part of the scene is illuminated. The reflected light is received by the image sensor. Reflected light from the scene is passed through the free-space optical mixer to downconvert the megahertz level amplitude modulation to hertz level beat tones which falls within the bandwidth of the image sensor. \textbf{b} The flash lidar system consists of a light source, an optical mixer, and an image sensor.}
\label{Big_Picture}
\end{center}
\end{figure*}

State of the art phase-shift based ToF imaging sensors rely on the photonic mixer device (PMD)~\cite{pmd}. Megahertz modulation frequencies are measured by electronic demodulation inside every pixel. These pixels are referred to as demodulation pixels~\cite{demodulation_pixel}. Homodyne detection is usually used to sample four different phases for the illumination. Since phase is measured, there is an ambiguity in the distance when a single frequency is used, and there is a trade-off between unambiguous range and depth resolution due to the amplitude modulation frequency selected. To significantly improve the unambiguous range while retaining the depth resolution, the phase of light at multiple amplitude modulation frequencies can be measured, and signal processing techniques similar to SFCW radar can be used. 

The ToF camera using PMD technology or similar architectures use an image sensor with specialized pixels, and therefore have limited spatial resolution. Since these systems use non-standard image sensors, they are expensive. Additionally, detecting multiple frequencies simultaneously requires multi-heterodyne detection, and this requires increasingly complex "smart pixels" with large sizes, leading to large image pixels and therefore reducing spatial resolution. Standard ToF cameras measure the phase at each frequency by stepping the frequency and measuring the phase, increasing the measurement time~\cite{ToF_frequency_stepping}.

One common problem with flash lidars is multi-path interference (MPI). Light might bounce several times in the scene and arrive at an image sensor pixel via different paths, corrupting phase estimates and therefore distance estimates. MPI is especially a big problem if there are highly reflective objects (specular or shiny) in the scene. There are solutions to overcome MPI. One approach is to use multiple frequencies to remove MPI in the scene rather than extending the unambiguous range~\cite{MPI_freq_1,MPI_freq_2}. The multiple frequencies, however, could still be used to extend the unambiguous range by correcting for MPI using other methods~\cite{MPI_without_freq}. In the rest of this paper, we neglect MPI effects and assume they are corrected or have minimal impact on measurements, and therefore we use the available frequency support for unambiguous range extension. 

\begin{figure}[t!]
\begin{center}
\includegraphics[width=1\textwidth]{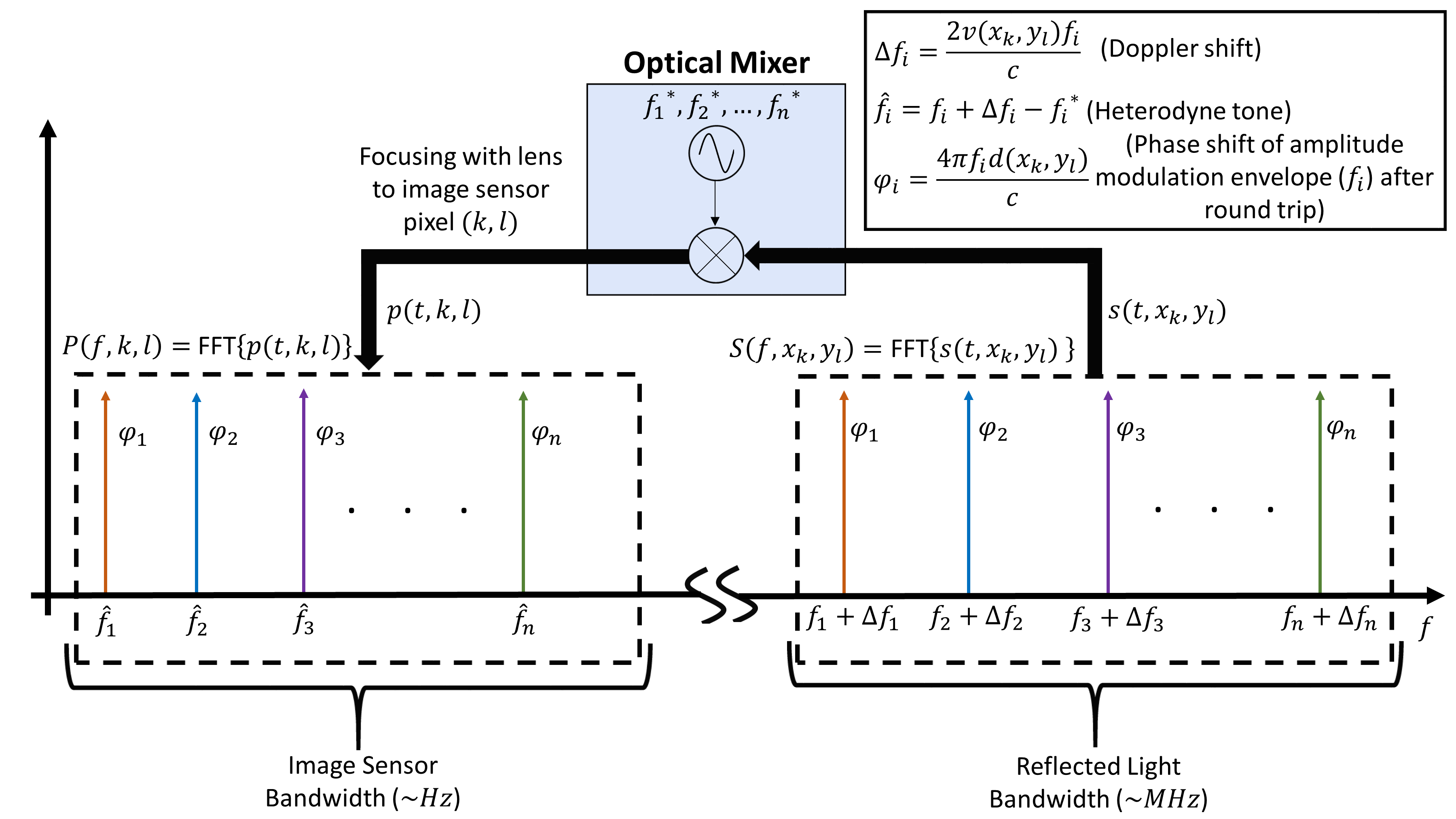}
\caption{The optical mixer bridges the gap between the megahertz frequencies which carry the phase and Doppler information from the scene per pixel to the hertz level bandwidth of the image sensor. The amplitude modulated light $s(t,x_k,y_l)$ is reflected from scene location $(x_k,y_l)$. Light downconverted by the optical mixer $p(t,k,l)$ is detected by image sensor pixel $(k,l)$ corresponding to the location $(x_k,y_l)$ in the scene.}
\label{Optical_Mixer}
\end{center}
\end{figure}

One possible way of measuring the phase of the incoming light modulated at megahertz frequency with a standard image sensor per pixel is by using an optical mixer (also referred to as an optical shutter) in front of the sensor to downconvert the high frequency to a lower beat tone (heterodyne detection). The system level architecture of the ToF imaging system is demonstrated in Figure~\ref{Big_Picture}, which shows the three main components of the ToF imaging system: modulated light source, free-space optical mixer, and the CMOS image sensor. Such an architecture would allow the use of the most advanced state of the art image sensors, which are low cost and have high spatial resolution. Such an architecture, however, ideally requires a free-space optical mixer with wide acceptance angle, low cost, low power consumption, and centimeter level aperture to be placed in front of the image sensor for performing the heterodyne detection. The function of the optical mixer is shown in Figure~\ref{Optical_Mixer}, in which the megahertz level amplitude modulated light reflected from the scene is downconverted by the optical mixer to hertz level beat tones. This allows the image sensor to detect the beat tones, which are used to estimate distance and velocity in the scene using signal processing techniques.

There have been previous attempts in designing a free-space optical mixer, however, all of these approaches have one or more drawbacks. A mechanical shutter is not practical since megahertz modulation frequencies requires extremely high rotation speeds, and this method has reliability issues due to moving parts. An image intensifier can be used for demodulation~\cite{image_intensifier_1,image_intensifier_2,image_intensifier_3}, however, the image intensifier is large in size and requires high operating voltages. Pockels cell sandwiched between polarizers can be used, but Pockels cells with centimeter level apertures are large and have prohibitively high half-wave modulation voltages~\cite{pockels}. Electro-absorption in multiple quantum well using an optical cavity can be used to modulate light\cite{stark_1}, but this approach has a narrow acceptance angle for light due to the use of an optical cavity in the modulator. Stepped quantum well modulator (SQM) has also been used to modulate light, but this design has limited aperture ($\sim$1~mm) and uses a microscope objective to focus the received light from the scene onto the surface of the  SQM\cite{stark_2}.

To design a free-space optical mixer with low half-wave modulation voltage, a resonant device is required. We avoid using an optical cavity since an optical cavity has a narrow acceptance angle for light, so we instead use an acoustic cavity.

In this paper, the design of an optical mixer relying on the photoelastic effect is demonstrated. The photoelastic modulator is a Y-cut lithium niobate wafer and is used to modulate the polarization state of light. Sandwiching the photoelastic modulator between two polarizers comprises the free-space optical mixer, which converts polarization modulation into intensity modulation. The optical mixer can be used with a standard CMOS image sensor to measure distances and velocity in a scene by flashing the scene with an amplitude modulated light source. 

\section{System Overview}
In this paper, we demonstrate the working principle of a prototype phase-shift based ToF imaging system with a standard CMOS image sensor using a resonant photoelastic modulator. A part of a scene is illuminated with amplitude modulated light and the reflected light from the scene is downconverted by an optical mixer and then imaged on a CMOS image sensor. The optical mixer consists of a photoelastic modulator sandwiched between polarizers. The photoelastic modulator is a 0.5~mm thick and 5.08~cm diameter Y-cut lithium niobate wafer with longitudinal and transparent electrodes. The photoelastic modulator modulates the polarization of light by operating the lithium niobate wafer at its  mechanical resonance modes. To demonstrate proof of concept, light of wavelength 630~nm is amplitude modulated at two frequencies and downconverted by the optical mixer such that the two beat tones fall within the bandwidth of the image sensor. We demonstrate the detection of two beat tones using heterodyne detection with a CMOS image sensor. This opens the way for simultaneous multi-frequency operation which can play a critical role as a flash lidar for various applications.

\section{Polarization Modulation by Photoelastic Effect}
In this section, the applied voltage to the photoelastic modulator will be related to the change in the polarization state of light passing through the modulator. The polarization modulation will be determined by calculating the modulated index ellipsoid for the photoelastic modulator. 

The index ellipsoid determines how light propagates in a material. The index ellipsoid can be modulated by using the photoelastic effect. Using the piezoelectric effect, strain can be generated in a wafer to control the polarization state of light electronically by modulating the index ellipsoid. The polarization modulation should be such that the two in-plane refractive indices for the wafer are modulated by different amounts to result in an in-plane polarization rotation for light.

\begin{figure*}[h]
\begin{center}
\includegraphics[width=1\textwidth]{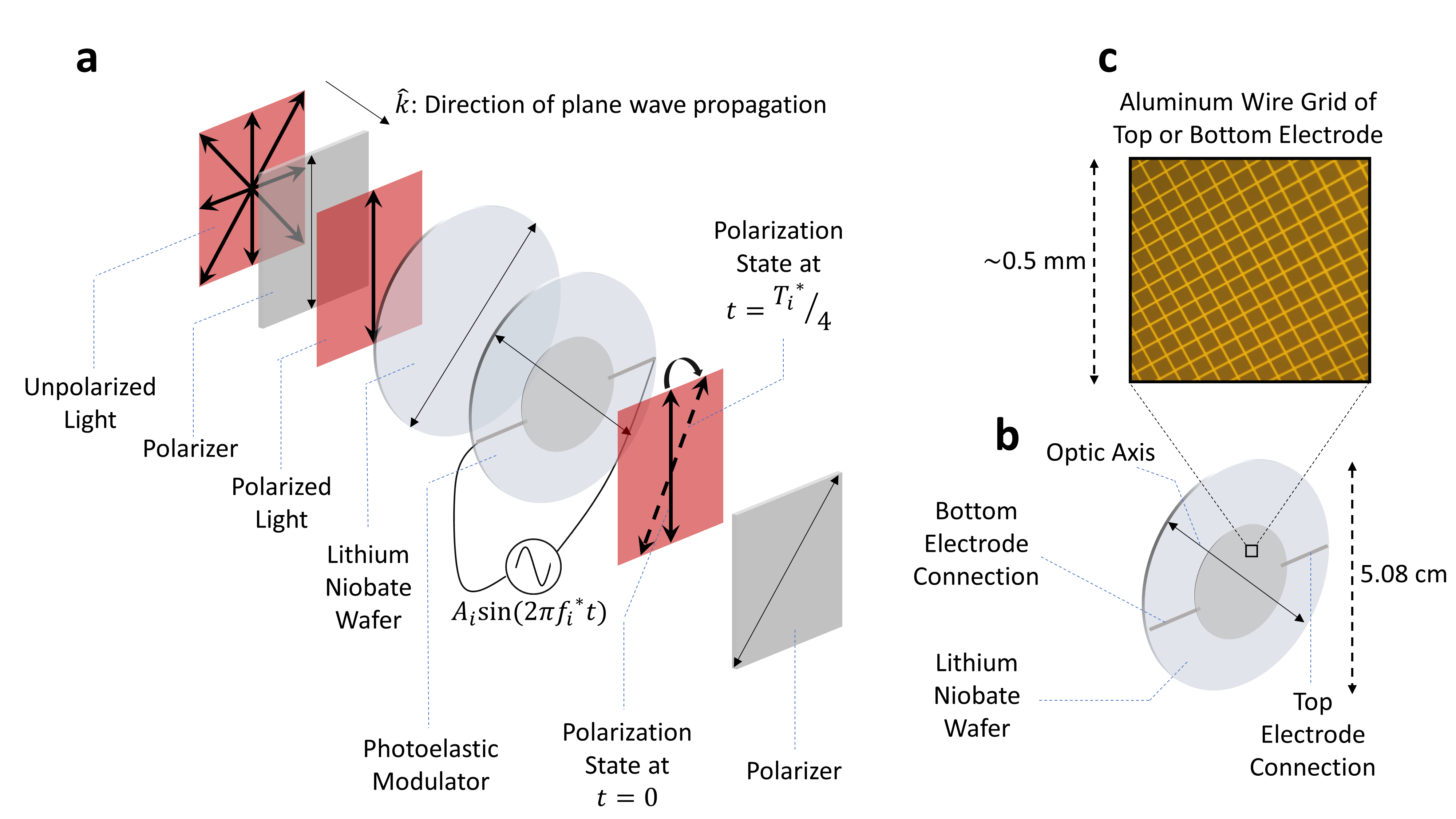}
\caption{\textbf{Polarization modulation using the photoelastic effect.} \textbf{a} The reflected light from the scene is polarized after passing through the polarizer. The photoelastic effect is used to rotate the polarization state of light in the lithium niobate wafer. The polarization of light is rotated at the applied tone $f_i^*$ (corresponding to period $T_i^*$) to the lithium niobate wafer. The polarization state of light is demonstrated at two different times. \textbf{b} The photoelastic modulator consists of a Y-cut lithium niobate wafer with longitudinal (perpendicular to the incoming light direction) and transparent electrodes. The longitudinal electrodes are aluminum wire grid deposited on the front and back surfaces (aligned) of the lithium niobate wafer. \textbf{c} A zoomed in version of the aluminum wire grid deposited on the wafer is shown. Each aluminum wire is 4~$\mu$m thick and the wires are separated by 40~$\mu$m, allowing optical transparency for the electrodes.}
\label{Polarization_Figure}
\end{center}
\end{figure*}

Photoelastic modulators are used commercially to control the polarization state of light, but they generally use a non-piezoelectric and isotropic material with transverse (parallel or nearly parallel to the incoming light direction) piezoelectric transducers to generate strain in the sample~\cite{photoelastic_transverse_1,photoelastic_transverse_2}. This configuration automatically breaks in-plane symmetry and leads to in-plane polarization modulation. The fundamental mechanical resonance frequencies for these devices are usually in the kilohertz range due to the centimeter scale optical aperture. Higher order mechanical modes can be used to drive the modulator, but as the mode order increases, the volume average for strain in the sample decreases due to the varying sign of the strain in the sample. Therefore, using transverse electrodes for the photoelastic modulator limits the mechanical resonance frequencies to kilohertz range, greatly limiting the depth resolution of an imaging system. To achieve megahertz mechanical resonance frequencies and square-centimeter-level apertures with high modulation efficiency, the electrodes need to be placed normal to the incoming light direction. If a standard wafer of thickness 0.5~mm is used, the fundamental mechanical resonance frequency will appear at roughly 4~MHz for lithium niobate, with resonance frequencies reaching up to 100~MHz (although as the mode order increases, the modulation efficiency drops).

If an isotropic material is used for polarization modulation, applying strain in the longitudinal direction (normal to the wafer) does not result in a change in the in-plane refractive indices due to in-plane symmetry with respect to the excitation. We therefore use a Y-cut lithium niobate wafer as the photoelastic modulator, breaking in-plane symmetry and leading to a net polarization modulation when longitudinal electrodes are used to generate strain in the wafer. 

Lithium niobate and many other piezoelectric materials are birefringent. Using a birefringent wafer leads to a static polarization rotation, which is different for rays incident on the wafer at different angles. Not correcting for this static birefringence will lead to a limited acceptance angle for the wafer. To correct for this static birefringence, which is standard practice in the design of wave plates, another identical wafer is placed parallel to the original wafer but rotated in plane by 90$^\circ$. Figure~\ref{Polarization_Figure} demonstrates the polarization modulation by the photoelastic modulator.

\begin{figure*}[h]
\begin{center}
\includegraphics[width=0.8\textwidth]{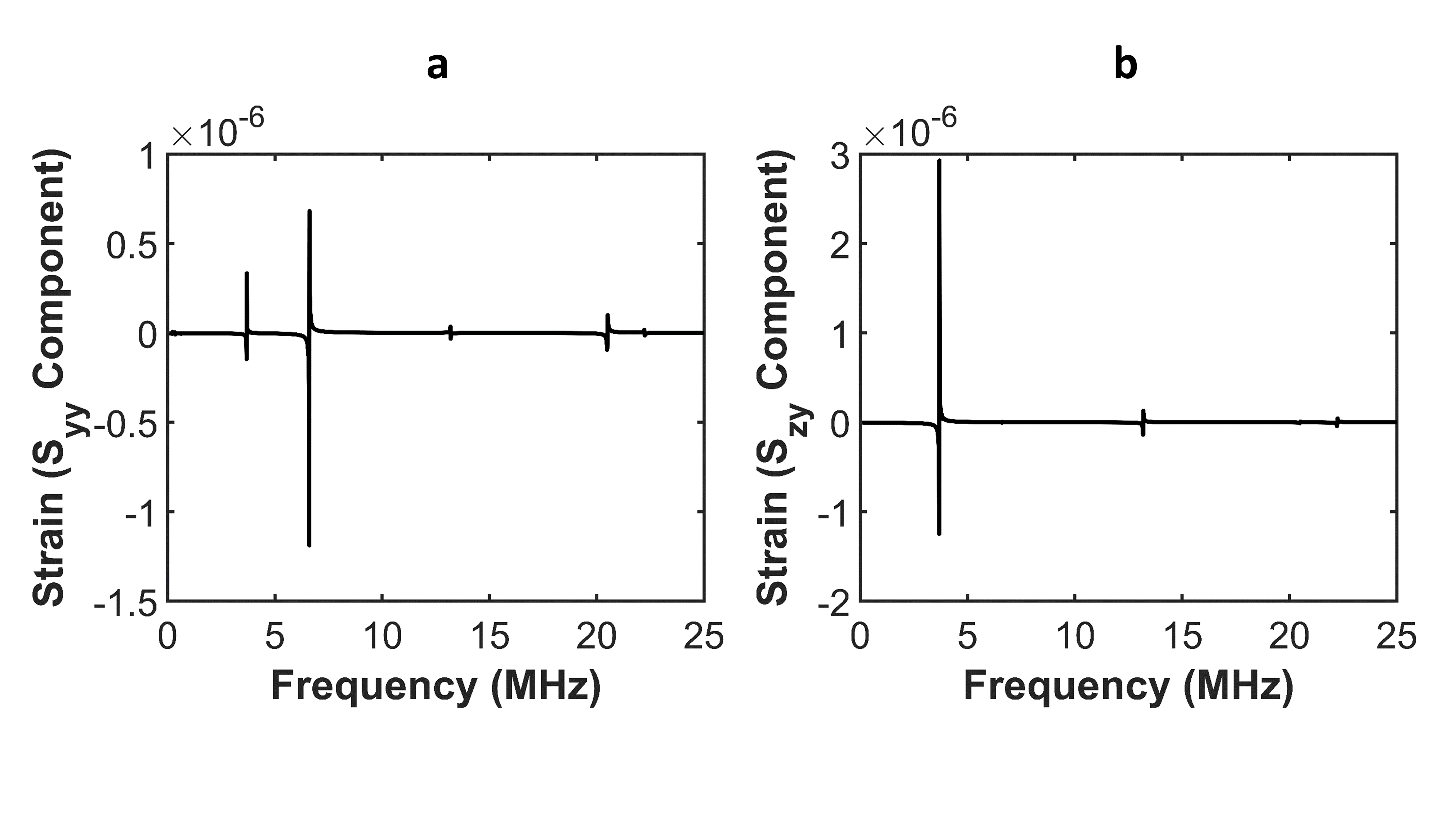}
\caption{\textbf{Strain volume average of the wafer swept from (0.1-25)~MHz with frequency stepping of 10~kHz.} The volume average for strain components which contribute to polarization modulation are plotted. \textbf{a}  $\bar{S}_{yy}$ strain component plotted against frequency. \textbf{b}  $\bar{S}_{zy}$ strain component plotted against frequency.}
\label{1-100MHz_Complete}
\end{center}
\end{figure*}

\begin{figure*}[h]
\begin{center}
\includegraphics[width=1\textwidth]{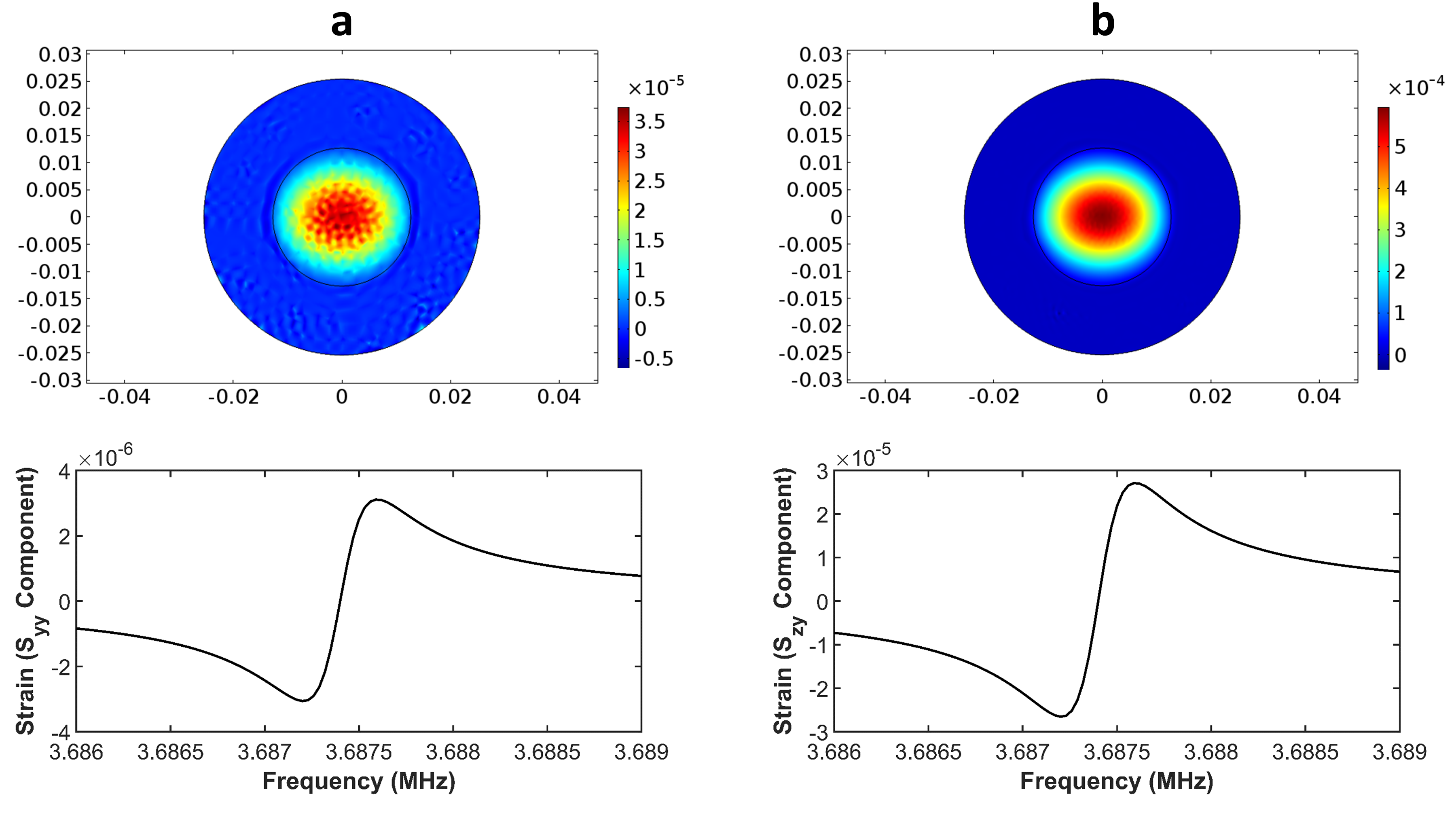}
\caption{\textbf{The strain profile of the wafer around its fundamental mechanical resonance frequency at 3.7~MHz with 2~V peak-to-peak applied through the electrodes and the wafer having a mechanical quality factor (Q) of around 9000}. \textbf{a} The bottom plot is the volume average for the $S_{yy}$ strain component in the wafer around the fundamental mechanical resonance frequency. The upper plot shows the cross-sectional $S_{yy}$ strain component of the wafer at the center of the wafer at resonance. \textbf{b} The bottom plot is the volume average for the $S_{zy}$ strain component in the wafer around the fundamental mechanical resonance frequency. The upper plot shows the cross-sectional $S_{zy}$ strain component of the wafer at the center of the wafer at resonance.}
\label{Fundamental_Mode}
\end{center}
\end{figure*}

If the strain profile is uniform or nearly uniform across the cross section of the wafer, to first order a single index ellipsoid can be used to describe the polarization modulation of light as it passes through the wafer. This approximation will be used throughout this section. The unmodulated index ellipsoid for the lithium niobate wafer can be written as in \eqref{Eq.1}, where $n_0$ and $n_e$ are the ordinary and extraordinary refractive indices of lithium niobate, respectively. 

\begin{gather}
\frac{x^2}{n_o^2} + \frac{y^2}{n_o^2} + \frac{z^2}{n_e^2} = 1 \label{Eq.1}
\end{gather}

To determine the effective index ellipsoid after strain is generated in the wafer through the piezoelectric effect, the wafer will be separated into infinitesimal volumes which have an infinitesimal thickness $dy$ along the y direction of the crystal and other dimensions equal to the wafer cross-section. Using the strain components, the polarization modulation can be determined for each of these infinitesimal volumes using the photoelastic effect. Let \textbf{S} denote the strain tensor in the wafer. The strain tensor is expressed as follows: $\textbf{S} = [S_{xx},S_{yy},S_{zz},2S_{zy},2S_{xz},2S_{yx}]$. The modulated index ellipsoid for this infinitesimal volume is expressed in \eqref{Eq.2}, where $p_{kl}$ are the photoelastic constants of lithium niobate for $(k,l) \in \{1,2,...,6\}$.

\begin{gather}
x^2\Big(\frac{1}{n_o^2} + p_{11}S_{xx} + p_{12}S_{yy}+ p_{13}S_{zz}+2p_{14}S_{zy}\Big) + y^2\Big(\frac{1}{n_o^2}+p_{12}S_{xx}+p_{11}S_{yy}+p_{13}S_{zz}-2p_{14}S_{zy}\Big) + \nonumber \\ z^2\Big(\frac{1}{n_e^2}+p_{13}S_{xx}+p_{13}S_{yy}+p_{33}S_{zz}\Big) + 2yz\Big(p_{41}S_{xx}-p_{41}S_{yy}+2p_{44}S_{zy}\Big) + \nonumber \\ 2zx\Big(2p_{44}S_{xz}+2p_{41}S_{yx}\Big) + 2xy\Big(2p_{14}S_{xz}+(p_{11}-p_{12})S_{yx}\Big) = 1 
\label{Eq.2}
\end{gather}

To first order, the effective index ellipsoid for the wafer is the arithmetic average of the index ellipsoids for these infinitesimal volumes. The effective index ellipsoid can be expressed as in \eqref{Eq.3}, where $\bar{S}_{ij}$ is the volume average for strain component in the wafer for $(i,j) \in \{x,y,z\}$.

\begin{gather}
x^2\Big(\frac{1}{n_o^2} + p_{11}\bar{S}_{xx} + p_{12}\bar{S}_{yy}+ p_{13}\bar{S}_{zz}+2p_{14}\bar{S}_{zy}\Big) + y^2\Big(\frac{1}{n_o^2}+p_{12}\bar{S}_{xx}+p_{11}\bar{S}_{yy}+p_{13}\bar{S}_{zz}-2p_{14}\bar{S}_{zy}\Big) + \nonumber \\ z^2\Big(\frac{1}{n_e^2}+p_{13}\bar{S}_{xx}+p_{13}\bar{S}_{yy}+p_{33}\bar{S}_{zz}\Big) + 2yz\Big(p_{41}\bar{S}_{xx}-p_{41}\bar{S}_{yy}+2p_{44}\bar{S}_{zy}\Big) + \nonumber \\ 2zx\Big(2p_{44}\bar{S}_{xz}+2p_{41}\bar{S}_{yx}\Big) + 2xy\Big(2p_{14}\bar{S}_{xz}+(p_{11}-p_{12})\bar{S}_{yx}\Big) = 1 
\label{Eq.3}
\end{gather}

We use the volume average of strain for the rest of the calculations. To determine the volume average strain tensor components generated in the lithium niobate wafer when voltage is applied through longitudinal electrodes, we simulate the wafer using the mechanics and piezoelectric modules in COMSOL~\cite{COMSOL5} simulation platform in frequency domain. 

The electrodes only cover half of the surface area for the wafer to limit clamping losses when the wafer is tested experimentally, as shown in Figure~\ref{Polarization_Figure}. For megahertz mechanical frequencies at room temperature, clamping losses are usually the dominant loss mechanism. The wafer will be clamped from the sides, therefore only the center part is deposited with aluminum wire grids and the light is passed through this section for polarization modulation.

The strain tensor components are calculated in the frequency domain from (0.1-25)~MHz with a frequency stepping of 10~kHz. Since the net polarization rotation of light is important, we calculate the volume average for the strain components. It is seen from COMSOL simulations that $S_{yy}$ and $S_{zy}$ with respect to crystal axis are the strain components which have a significant non-zero volume average for strain. The effective index ellipsoid can therefore be expressed as \eqref{Eq.4}. 

\begin{gather}
x^2\Big(\frac{1}{n_o^2} + p_{12}\bar{S}_{yy}+2p_{14}\bar{S}_{zy}\Big) + y^2\Big(\frac{1}{n_o^2}+p_{11}\bar{S}_{yy}-2p_{14}\bar{S}_{zy}\Big)  + \nonumber \\ z^2\Big(\frac{1}{n_e^2}+p_{13}\bar{S}_{yy}\Big) + 2yz\Big(-p_{41}\bar{S}_{yy}+2p_{44}\bar{S}_{zy}\Big) = 1 
\label{Eq.4}
\end{gather}

We apply a rotation to the yz axis such that the new form is diagonal~\cite{crystal_optics}. Using the coordinate transformations in \eqref{Eq.5}, \eqref{Eq.4} can be transformed into \eqref{Eq.6}.

\begin{gather}
y = y'\text{cos} \theta - z'\text{sin} \theta \nonumber \\ 
z = y'\text{sin} \theta + z'\text{cos} \theta \label{Eq.5}
\end{gather}

\begin{gather}
x^2\Big(\frac{1}{n_o^2} + p_{12}\bar{S}_{yy}+2p_{14}\bar{S}_{zy}\Big) + {y'}^2\Big(\frac{1}{n_o^2} + p_{11}\bar{S}_{yy}-2p_{14}\bar{S}_{zy}+(2p_{44}\bar{S}_{zy}-p_{41}\bar{S}_{yy})\text{tan} \theta\Big) + \nonumber \\ {z'}^2\Big(\frac{1}{n_e^2} + p_{13}\bar{S}_{yy} - (2p_{44}\bar{S}_{zy}-p_{41}\bar{S}_{yy})\text{tan} \theta\Big) = 1 \nonumber \\
\text{tan} (2 \theta) = \frac{4p_{44}\bar{S}_{zy} - 2p_{41}\bar{S}_{yy}}{\Big(\frac{1}{n_o^2}+p_{11}\bar{S}_{yy}-2p_{14}\bar{S}_{zy}\Big) - \Big(\frac{1}{n_e^2}+p_{13}\bar{S}_{yy}\Big)} \label{Eq.6}
\end{gather}


Since $\text{tan}\theta \ll 1$, we neglect the modulations of the $y'$ and $z'$ axis which include the $\text{tan}\theta$ term. We assume for our analysis that the beam is incident at an angle $\theta$ to the normal. Since $\theta$ < 1$^\circ$ usually, the path traversed by the beam is approximately equal to the thickness of the wafer.

\begin{gather}
x^2\Big(\frac{1}{n_o^2} + p_{12}\bar{S}_{yy}+2p_{14}\bar{S}_{zy}\Big) + {y'}^2\Big(\frac{1}{n_o^2}+p_{11}\bar{S}_{yy}-2p_{14}\bar{S}_{zy}\Big) + {z'}^2\Big(\frac{1}{n_e^2}+p_{13}\bar{S}_{yy}\Big) = 1 \label{Eq.7}
\end{gather}

Figure~\ref{1-100MHz_Complete} shows the volume average of the strain components in the wafer $\bar{S}_{yy}$ and $\bar{S}_{zy}$ corresponding to the region covered with longitudinal electrodes. We see resonances at multiple frequencies, but for the rest of this paper we will be focusing on the resonance frequencies at the fundamental mechanical resonance frequency for the wafer at roughly 3.7~MHz and the resonance frequency at roughly 20.5~MHz. We first consider the fundamental mode at 3.7~MHz. The cross section of the wafer at the center for the $S_{yy}$ and $S_{zy}$ strain components around the fundamental mechanical resonance frequency and the volume average for the strain components inside the wafer are shown in Figure~\ref{Fundamental_Mode}. When the wafer is driven at one of its mechanical resonance frequencies $f_i^*$, the volume average strain components can be expressed as $\bar{S}_{yy} = A_1\text{cos}(2 \pi f_i^*t)$ and $\bar{S}_{zy} = A_2\text{cos}(2 \pi f_i^*t)$. The modified index ellipsoid in this case can be expressed as in \eqref{Eq.8}.

\begin{gather}
x^2\Big(\frac{1}{n_o^2} + p_{12}A_1\text{cos}(2 \pi f_i^*t)+2p_{14}A_2\text{cos}(2 \pi f_i^*t)\Big) + {y'}^2\Big(\frac{1}{n_o^2}+p_{11}A_1\text{cos}(2 \pi f_i^*t)-2p_{14}A_2\text{cos}(2 \pi f_i^*t)\Big) + \nonumber \\ {z'}^2\Big(\frac{1}{n_e^2}+p_{13}A_1\text{cos}(2 \pi f_i^*t)\Big) = 1 \label{Eq.8}
\end{gather}

The electro-optic effect has negligible effect compared to the photoelastic effect due to the high mechanical resonance exhibited by the wafer, therefore the electro-optic effect will not be included in the polarization modulation calculations. In the next section, polarization modulation for an incoming beam along the $y'$ direction of the crystal will be calculated. It can be shown that the acceptance angle for this type of photoelastic modulator is roughly 20$^\circ$ due to birefringence of the wafer when 0.5~mm thick wafer is used along with another identical wafer placed parallel and rotated in plane by 90$^\circ$. A thinner wafer can be used to increase the acceptance angle (e.g. 0.1~mm). More detailed analysis for arbitrary angles, field of view, and taking the electro-optic effect into account will be explained in a future work. 

\subsection{Normal Incidence}
In this section, the polarization modulation $\Delta \theta(t)$ as a function of time will be derived assuming the incoming beam is perpendicular to the wafer (actually at an angle $\theta$ to the normal of the wafer) and the wafer is driven at its fundamental mechanical resonance frequency of $f_i^*$. Another identical wafer parallel and rotated in plane by 90$^\circ$ is placed after the photoelastic modulator to correct for static polarization rotation of light. The incoming beam sees the refractive indices $n_{z'}(t)$ and $n_x(t)$ when passing through the photoelastic modulator, where refractive index along the x and z directions are modulated by the photoelastic effect as in \eqref{Eq.8}. 

\begin{gather}
n_x(t) = \frac{n_0}{\sqrt{1+n^2_0(p_{12}A_1\text{cos}(2\pi f_i^*t)+2p_{14}A_2\text{cos}(2\pi f_i^*t))}} \label{Eq.9}
\end{gather}

Since $n^2_0(p_{12}A_1\text{cos}(2\pi f_i^*t)+2p_{14}A_2\text{cos}(2\pi f_i^*t)) \ll 1$, we can approximate \eqref{Eq.9} as shown in \eqref{Eq.10}. 

\begin{gather}
n_x(t) \approx n_0 - n^3_0\Big(\frac{1}{2}p_{12}A_1 + p_{14}A_2\Big)\text{cos}(2\pi f_i^*t) \label{Eq.10}
\end{gather}

\begin{gather}
n_{z'}(t) = \frac{n_e}{\sqrt{1+n^2_ep_{13}A_1\text{cos}(2 \pi f_i^*t)}} \label{Eq.11}
\end{gather}

\begin{gather}
n_{z'}(t) \approx n_e - \frac{1}{2}n^3_e p_{13}A_1\text{cos}(2\pi f_i^*t) \label{Eq.12}
\end{gather}

The change in the in-plane refractive indices $\Delta n(t)$ is expressed in \eqref{Eq.13}. 

\begin{gather}
\Delta n(t) = (n_0 - n_e) - (n_x(t) - n_{z'}(t))  = \bigg(n^3_0 \Big(\frac{1}{2}p_{12}A_1 + p_{14}A_2\Big) - \frac{1}{2}n^3_e p_{13}A_1\bigg)\text{cos}(2\pi f_i^*t) \label{Eq.13}
\end{gather}

The polarization change of light after passing through the wafer of thickness $d$ with wavelength of light $\lambda$ is expressed in \eqref{Eq.14}.

\begin{align}
\Delta \theta(t) = \frac{2 \pi}{\lambda}\Delta n(t)d \label{Eq.14}
\end{align}

\subsubsection{Depth of Polarization Modulation}
In this section, the relationship between the depth of polarization modulation $D$ as a function of the applied peak-to-peak voltage $V_p$ to the photoelastic modulator and the quality factor of the fundamental mechanical resonance mode $Q$ of the wafer will be derived. We calculate the depth of polarization modulation assuming normal incidence of light to the lithium niobate wafer at the fundamental mechanical resonance frequency for the wafer. We calculate the volume average for the two strain components ($S_{yy}$ and $S_{zy}$) contributing to polarization modulation in the sample using COMSOL. Loss is added to the lithium niobate wafer to determine the strain components and therefore the depth of modulation at a given mechanical quality factor and voltage applied to the electrodes. 

In simulation, we apply 2~V peak-to-peak to the electrodes at around the fundamental mechanical resonance frequency for the wafer (approximately 3.7~MHz). From COMSOL simulations in Figure~\ref{Fundamental_Mode}, we see that the volume average at the resonance is roughly $A_1=3\times10^{-6}$ and $A_2=2\times10^{-5}$. Using the photoelastic constants $p_{12} = 0.06$, $p_{13}=0.172$, and $p_{14} = -0.052$ from \cite{photoelastic_constants} with \eqref{Eq.13} and \eqref{Eq.14}, the depth of polarization modulation is calculated to be 0.0715~radians for light of wavelength 630~nm. The quality factor for the wafer in the simulation with the added loss is roughly 9000 (calculated based on 3dB cut-off points for the strain around the fundamental mechanical resonance frequency). Based on these results, the depth of polarization modulation $D$ for an incident beam along the y direction of the wafer can be calculated roughly as in \eqref{Eq.15} for light of wavelength 630~nm:

\begin{equation}
\frac{D}{2} = \bigg(n^3_0 \Big(\frac{1}{2}p_{12}A_1 + p_{14}A_2\Big) - \frac{1}{2}n^3_e p_{13}A_1\bigg)\frac{2 \pi}{\lambda}d \approx  4\times 10^{-6} Q V_p \label{Eq.15}
\end{equation}

The depth of modulation is independent of the wafer thickness to first order, since the electric field inside the wafer is inversely proportional to wafer thickness, however, this is compensated by the larger path traversed by the light when passing through the wafer. The acceptance angle for a 0.5~mm thick wafer is roughly 20$^\circ$ when the static polarization correcting wafer is also used. The acceptance angle is calculated by finding the largest incoming angle with respect to the wafer normal such that the static birefringence between the ordinary and extraordinary rays is 90$^\circ$. A thinner wafer can be used to improve the acceptance angle for the photoelastic modulator while retaining the same depth of polarization modulation. 

\section{Polarization Modulation Conversion to Intensity Modulation}
Polarization modulation can be converted into intensity modulation by sandwiching the photoelastic modulator between two polarizers. Malus' law governs the transmitted intensity of light after passing through a polarizer: the transmitted intensity of light after passing through a polarizer is scaled by cosine squared of the angle between the polarization direction of light and the transmission axis of the polarizer. Since standard polarizers have high extinction ratios, high modulation depth can be realized. 

When the lithium niobate wafer is driven near its resonance mode(s), the intensity modulation is a cosine inside a cosine (similar to frequency modulation). This expression can be expanded by the Jacobi-Anger expansion, causing an infinite number of equally spaced frequencies. For each amplitude modulation frequency, the fundamental tone is downconverted into the bandwidth of the image sensor, and the fundamental tone is used for signal processing; the other tones are low-pass filtered by the image sensor.

\begin{figure*}[h]
\begin{center}
\includegraphics[width=0.7\textwidth]{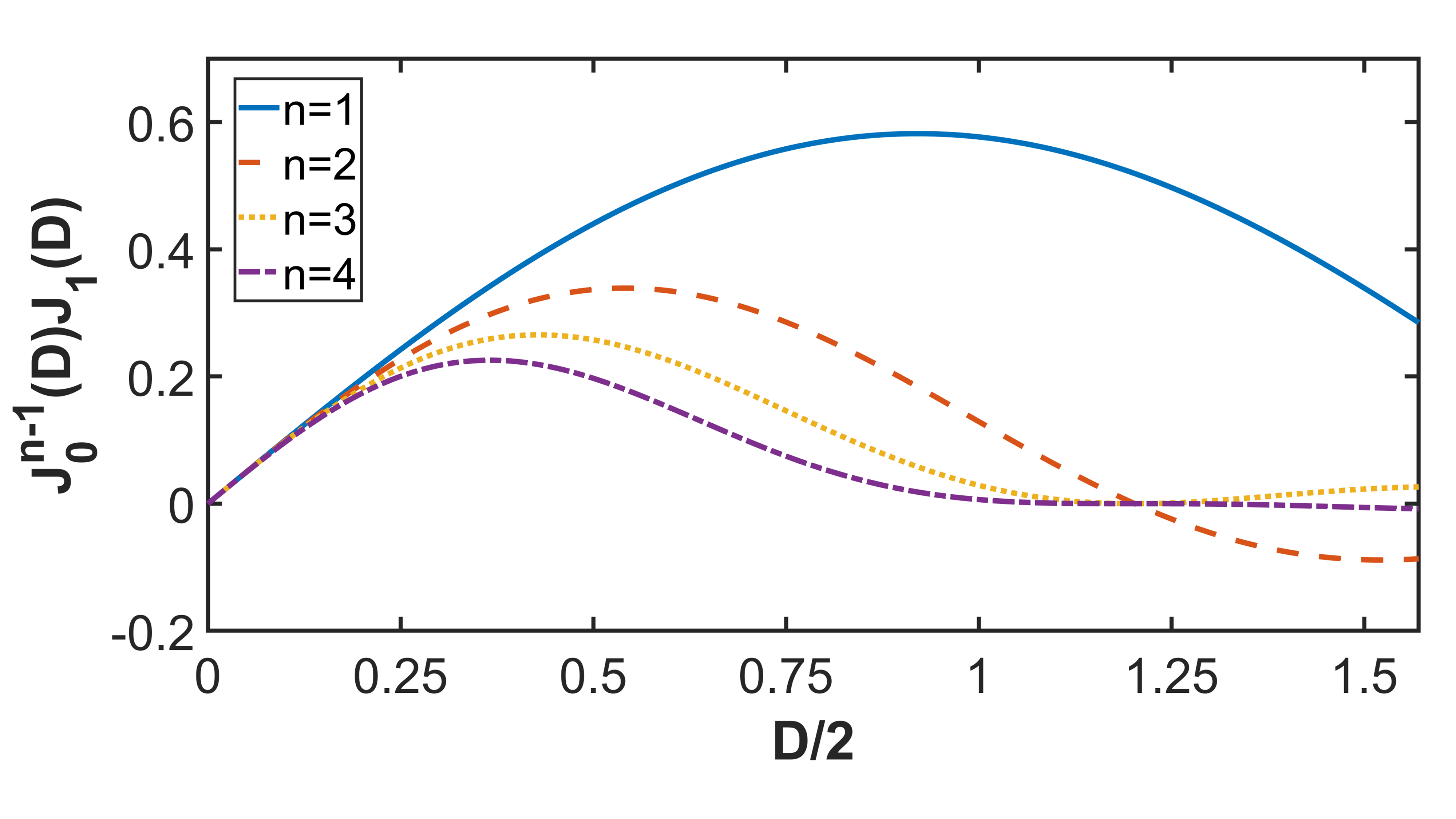}
\caption{The modulation depth of intensity as a function of phase modulation from the photoelastic modulator is demonstrated. As the number of frequencies used $n$ simultaneously increases, the attainable modulation depth drops for all frequencies.}
\label{Bessel_Optimum}
\end{center}
\centering
\end{figure*}

The scene is illuminated with intensity modulated light $I(t)$ at frequencies $f_1,f_2 ,..., f_n$ slightly detuned from the frequencies used to drive the photoelastic modulator $f_1^*,f_1^*,...f_n^*$. The light reflected from location $(x_l,y_l)$ with reflectivity $r(x_k,y_l)$ in the scene is represented as $s(t,x_k,y_l)$, which is Doppler shifted by $\Delta f_1,\Delta f_2 ,...., \Delta f_n$ and phase shifted by $\psi_1,\psi_2,..., \psi_n$. The received heterodyne beat signal $\hat{f}_i$ at image sensor pixel $(k,l)$ corresponding to scene location $(x_k,y_l)$ is represented as $p(t,k,l)$, which carries the phase and Doppler information at a frequency ($\sim$~Hz) which falls within the bandwidth of the image sensor. $p(t,k,l)$  represents the multiple beat frequencies detected by a single image sensor pixel, where $\theta(t)$ is the angle between the polarization direction of light that has passed through the photoelastic modulator and the second polarizer transmission axis. $\psi_i = \frac{4\pi f_i}{c}d(x_k,y_l)$ is the phase shift at the receiver of the amplitude modulated light that illuminates the scene, where $d(x_k,y_l)$ is the distance of the receiver to the scene location $(x_k,y_l)$. $\Delta f_i = \frac{2v_(x_k,y_l) f_i}{c}$ is the Doppler shift for the received light due to motion with velocity $v(x_k,y_l)$ in the scene location $(x_k,y_l)$. The distance and velocity of each point in the scene can be efficiently computed by performing a fast Fourier transform (FFT) with respect to time per image sensor pixel and using the phase and frequency shift information.

\begin{equation}
I(t) = I_0\sum_{i=1}^n \text{cos}(2\pi f_i t) \label{Eq.16}
\end{equation}

\begin{equation}
s(t,x_k,y_l) = r(x_k,y_l)I_0\sum_{i=1}^n\text{cos}(2\pi (f_i + \Delta f_i) t + \psi_i) \label{Eq.17}
\end{equation}

\begin{equation}
p(t,k,l) = \frac{1}{2}s(t,x_k,y_l)\underbrace{\text{cos}^2(\theta(t))}_\text{Malus' Law} \label{Eq.18} 
\end{equation}

\begin{equation}
\theta(t) = \underbrace{\frac{D}{2}\sum_{i=1}^n\text{cos}(2 \pi f_i^*t) }_\text{Phase Modulation from Modulator}  + \theta_0 \label{Eq.19}
\end{equation}

Due to optical mixing, which takes place in \eqref{Eq.18}, many tones are generated. The beat term which falls within the bandwidth of the image sensor is shown in \eqref{Eq.20}, with $\hat{f}_i~=~ f_i~+~\Delta f_i~-~f_i^*$.

\begin{align}
p(t,k,l) \approx -\frac{1}{4}I_0r(x_k,y_l){J_0}^{n-1}(D)J_1(D)\text{sin}(2\theta_0)\sum_{i=1}^n \text{cos}(2 \pi \hat{f}_it + \psi_i) \label{Eq.20}
\end{align}

Optimum depth of modulation can be calculated by optimizing ${J_0}^{n-1}(D)J_1(D)\text{sin}(2\theta_0)$, assuming the depth of polarization modulation $D$ is the same for all frequencies $f_i^*$. The sin term is maximized when $\theta_0=45^\circ$, suggesting that the angle between the two polarizer transmission axis should have a 45$^\circ$ angle difference. Figure~\ref{Bessel_Optimum} shows the depth of modulation for different $D$ values and number of frequencies $n$ used to drive the photoelastic modulator. The intensity modulation depth drops as the number of frequencies used to drive the photoelastic modulator is increased. Since the wafer is multi-moded and multiple of these modes are driven with the source, many mixing terms appear in the spectrum, reducing the depth of modulation. Additionally, the polarization modulation depth for higher order mechanical modes will be smaller compared to lower order modes. An alternative to using a single wafer driven at multiple of its resonance frequencies is to have wafers of different thicknesses that are placed in front and parallel to each other. Each wafer can then be driven at its fundamental mechanical resonance frequency, or possibly by driving another higher order mode.

\section{Multi-Frequency Operation to Extend Unambiguous Range}
Using a single frequency for distance measurements limits the unambiguous range or the depth resolution. When using a single frequency, which refers to the amplitude modulation frequency, the unambiguous range is limited to half the wavelength $\Big(\frac{c}{2f_i}\Big)$ corresponding to the frequency used. Using a low frequency results in a large unambiguous range, but the estimated phase needs to be accurate, since the calculated distance is directly proportional to the measured phase. Even small phase errors due to shot-noise or electronic noise will lead to significant distance errors, which necessitates using megahertz frequencies. If a single frequency is used, and the range is limited to $0 \leq d(x_k,y_l) < \frac{c}{2f_i}$, the measured phase for the beat tone $\psi_i \in [0,2 \pi)$ is used as in \eqref{Eq.21} to estimate the distance $d^*(x_k,y_l)$ corresponding to image sensor pixel $(k,l)$:

\begin{equation}
d^*(x_k,y_l) = \frac{\psi_ic}{4\pi f_i} \label{Eq.21}
\end{equation}

To significantly improve the unambiguous range while retaining the depth resolution, the phase of multiple frequencies can be used after the round-trip of light, similar to the operation of SFCW radar. The standard image sensor has high angular resolution and most of the light from the scene is reflected once, therefore the limitations that apply to SFCW radar do not apply. The high angular resolution provided by the image sensor limit the number of reflectors in the scene per sensor pixel to one. This allows achieving high depth resolution and unambiguous range despite measuring the phase of the returned light at several discrete frequencies.

There are two problems that need to be addressed: the number of  modulation frequencies to be used, and the reconstruction algorithm for estimating the distance and velocity per image sensor pixel. In this paper, we focus on the reconstruction algorithm for distance, and leave the selection of the modulation frequencies and velocity estimation as future work.

\begin{figure*}[h]
\begin{center}
\includegraphics[width=1\textwidth]{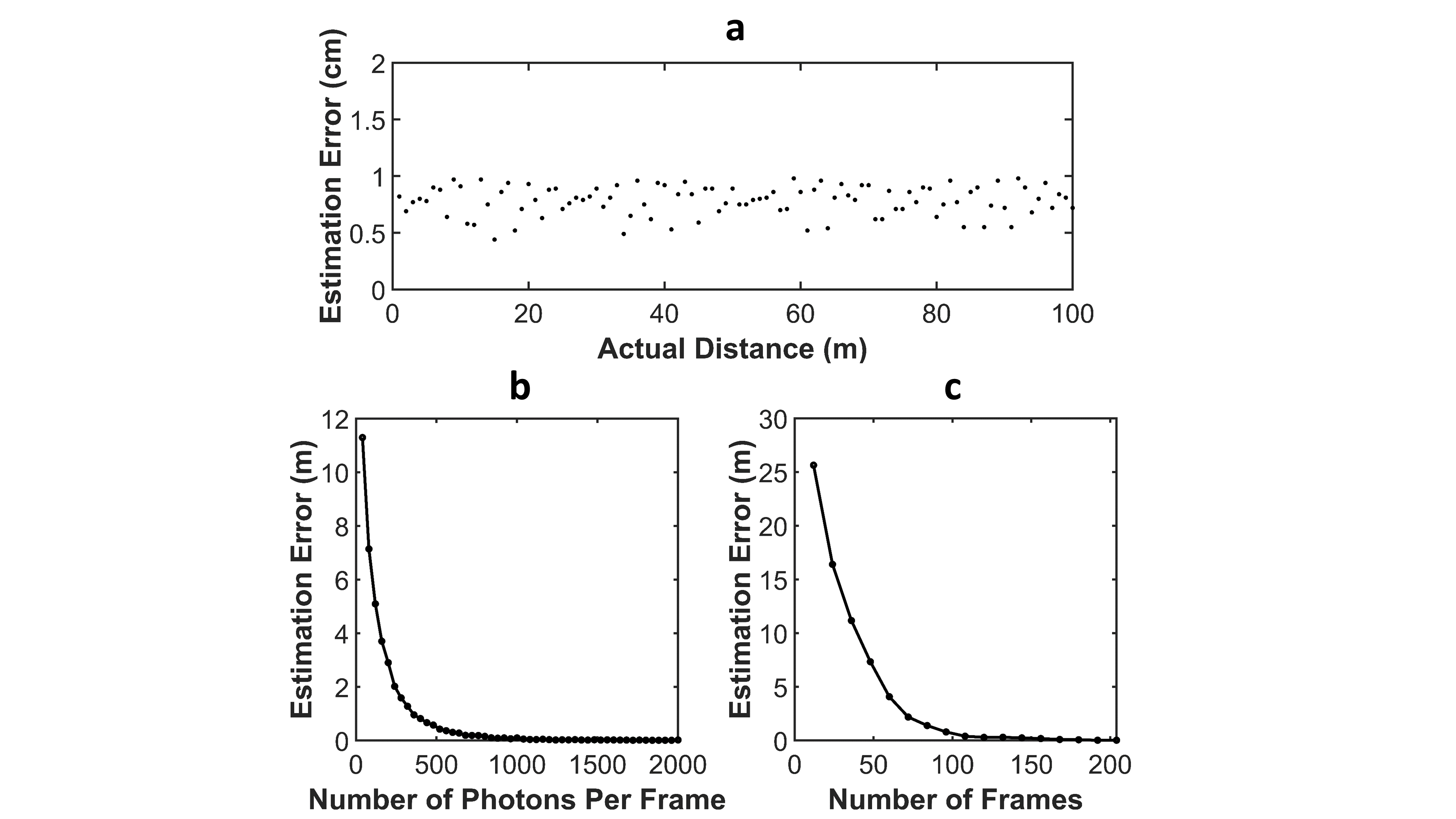}
\caption{\textbf{Simulation results: Distance estimation error using the forward reconstruction algorithm to optimize equation~\eqref{Eq.23}}. The modulation frequencies are 97.8~MHz, 19.59~MHz, and 4.02~MHz, with corresponding beat frequencies appearing at 80~Hz, 170~Hz, and 250~Hz, respectively. The distance is estimated from 1~m to 100~m with 1~m stepping. The assumed resolution is 1~cm for the forward reconstruction algorithm and 100 averages are used for error estimation. A sampling rate of 600~Hz is assumed for the image sensor. \textbf{a} 200 frames are captured (3Hz refresh rate for distance estimate). In each frame, each pixel captures 2000 photons and the measurement is shot-noise limited. The absolute value for the average distance error is about 0.8~cm. \textbf{b} 200 frames are captured, with shot-noise limited measurements. The estimation error averaged over (1-100)~m is plotted against the number of photons per pixel per frame. \textbf{c} Each pixel captures 2000 photons per frame, the measurements are shot-noise limited. The estimation error averaged over (1-100)~m is plotted against the number of frames used.} 
\label{cvx_performance}
\end{center}
\end{figure*}

We first solve the problem of finding an algorithm for distance reconstruction per sensor pixel $(k,l)$ corresponding to location $(x_k,y_l)$ in the scene, assuming modulation frequencies $f_1,...,f_n$ are used for illumination, and the phase response measured at each frequency using the optical mixer and an image sensor. Maximum likelihood detection is used for distance reconstruction per image sensor pixel to maximize the probability of correct detection. 

Before using the forward reconstruction algorithm for estimating the distance, we need to accurately predict the phase of each frequency sampled by the image sensor. This is equivalent to estimating the complex gains of a noisy mixture of sinusoids, where the noise is white and follows a Gaussian distribution. The phases for the mixture of noisy sinusoids can be estimated efficiently via the Newtonized orthogonal matching pursuit (NOMP)~\cite{nomp}. Once the phases have been extracted, each phase can be modeled as a Gaussian distribution: $\psi_i^* \sim \mathcal{N}\Big(\frac{4\pi d(x_k,y_l) f_i}{c} \text{ (mod } \textbf{ }2\pi),\sigma^2 \Big)$, with $d(x_k,y_l)$ the distance of the location in the scene $(x_k,y_l)$ to the receiver, $c$ the speed of light in the scene, and $\sigma^2$ the noise variance. Due to the $2\pi$ phase wrapping, even if multiple frequencies are used and perfect phase information is retrieved, there will always be an ambiguous range at the least common multiple of the wavelengths corresponding to the modulation frequencies. This presents an ill-posed optimization problem due to multiple solutions. As a way around this problem, we define an unambiguous range, which is smaller than the least common multiple of the modulation frequencies. In fact, this unambiguous range should be determined based on the signal-to-noise ratio (SNR) and the modulation frequencies, but that problem will not be dealt in this paper. 

We cast the distance estimation as an optimization problem, in which the most likely distance $d^*(x_k,y_l)$ to explain the observed phases within the selected unambiguous range is chosen as the distance estimate per image sensor pixel. If $\psi_i^*$ is the estimated phase corresponding to amplitude modulation frequency $f_i$, $d_u$ the selected unambiguous range, and $p_i$ the probability density function of a Gaussian random variable, the optimization problem can be expressed as in \eqref{Eq.22}.

\begin{equation}
\begin{aligned}
& \underset{d(x_k,y_l)}{\text{arg max}}
& & \prod_{i=1}^n p_i \bigg(\psi_i^* = \frac{4\pi d(x_k,y_l)f_i}{c} \text{ (mod } \textbf{ }2\pi) \bigg)  \\
& \text{subject to}
& & 0 \leq d \leq d_u \label{Eq.22}
\end{aligned}
\end{equation}

This is a non-convex optimization problem due to phase wrapping. One possible approach to solve the optimization problem is by separating the optimization problem into bounded least-squares problems through constraining the distance such that within each of the regions, the objective function is convex (possibly also with some approximations). The global maximum among the local maxima would then be equivalent to solving the non-convex optimization problem. We leave this approach as future work, and use a reconstruction algorithm based on forward reconstruction.

Taking the logarithm of \eqref{Eq.22}, this problem is equivalent to \eqref{Eq.23}, where $\mathbf{k}$ is a vector of integers to deal with phase wrapping.

\begin{equation}
\begin{aligned}
& \underset{d(x_k,y_l),\mathbf{k}}{\text{arg min}}
& & \sum_{i=1}^n \Big (\psi^*_i - \frac{4\pi d(x_k,y_l)f_i}{c}+2\pi \mathbf{k}(i)\Big)^2  \\
& \text{subject to}
& & 0 \leq d \leq d_u \text{, } \mathbf{k}(i) \in \{-1,0,1\} \label{Eq.23}
\end{aligned}
\end{equation}

We use forward reconstruction to estimate the distance $d^*(x_k,y_l)$ corresponding to image sensor pixel $(k,l)$. Within the unambiguous range $d_u$, we discretize the region $[0,d_u)$ with resolution $r$. We evaluate the phase that would have been observed if there was no noise corrupting the measurements for each frequency $f_i$ with $\psi_i = \frac{4\pi d_m(x_k,y_l) f_i}{c}$, where $d_m(x_k,y_l) = r\times m$, $m \in \mathbb{N}$. The distance is estimated by minimizing the objective function in \eqref{Eq.23}, and this procedure is applied for each image sensor pixel $(k,l)$ to estimate the distance in the scene $d^*(x_k,y_l)$.

We simulate the performance of the distance estimation algorithm per image sensor pixel assuming an unambiguous range of 100~m, camera frame rate of 600~Hz, shot-noise limited measurements with 3 modulation frequencies used at (97.8, 19.59, 4.02)~MHz and beat tones appearing at (80, 170, 250)~Hz, respectively. The performance of the algorithm for these parameters and as a function of number of frames and the number of photons per frame per pixel is shown in Figure~\ref{cvx_performance}. The average estimation error in the range (1-100)~m using 2000 photons per pixel per frame and 200 frames used per distance estimate is around 0.8~cm. Velocity estimation in a scene is not considered in this paper, but essentially the Doppler shift of the tones are used. The details for the estimation algorithm, choosing the frequencies to maximize depth resolution, unambiguous range, and extracting velocity from the scene will be explained in a future work. 

\section{Experiment}
\begin{figure}[H]
\begin{center}
\includegraphics[width=0.7\textwidth]{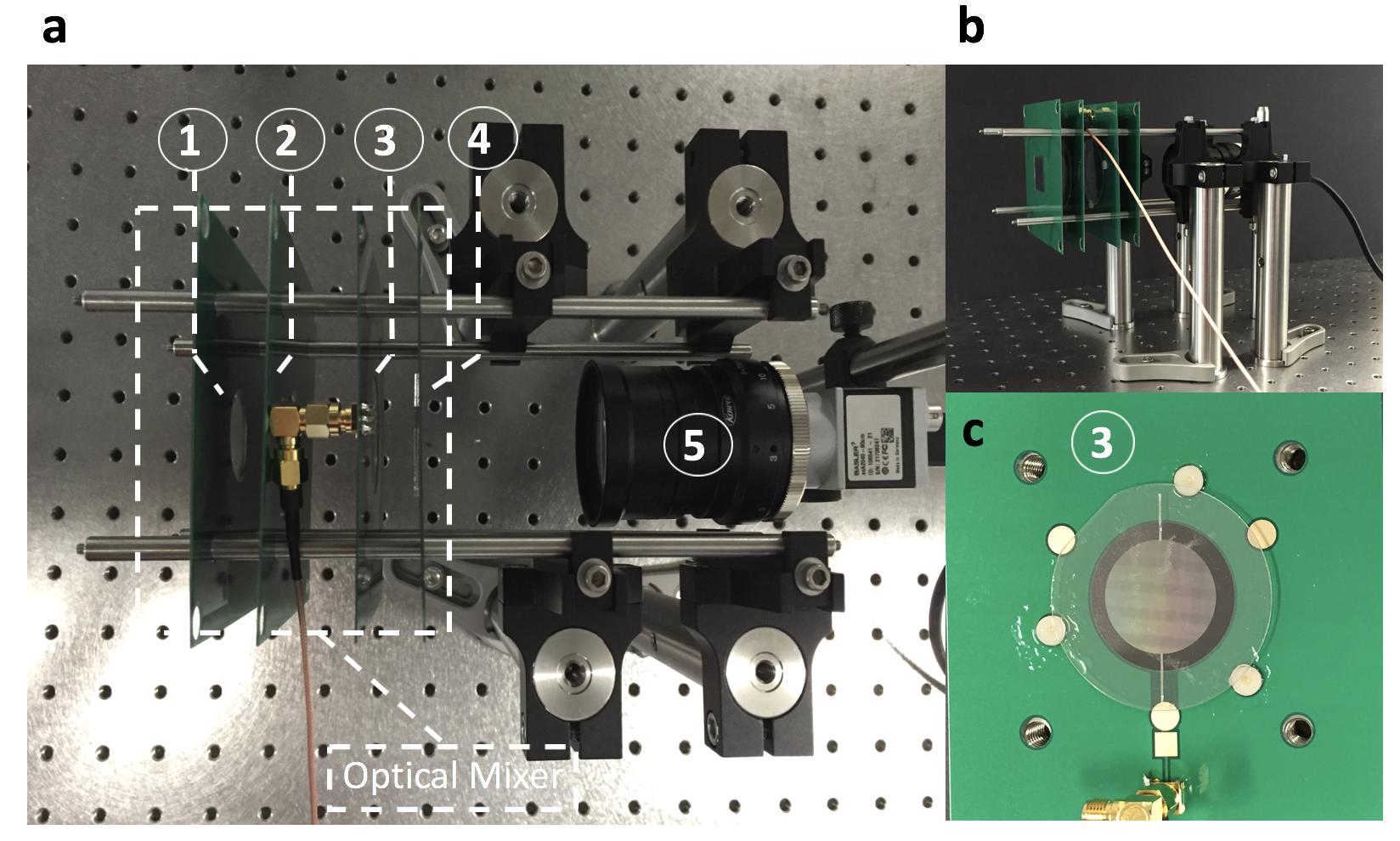}
\caption{\textbf{Prototype ToF imaging system.} \textbf{a} Top view of the ToF imaging system, including the optical mixer placed in front of a Basler Ace 2040-90um CMOS image sensor. 1: Polarizer, 2: Lithium niobate wafer, 3: Photoelastic modulator, 4: Polarizer, 5: CMOS Camera. \textbf{b} Side-view of ToF imaging system. \textbf{c} Photoelastic modulator: 0.5~mm thick and 5.08~cm diameter Y-cut lithium niobate wafer is coated on both sides with aluminum wire grid. The aluminum strips to the sides are for wirebonding to PCB to apply voltage to the modulator.}
\label{Measurement_Setup}
\end{center}
\end{figure}

A Y-cut lithium niobate wafer of 0.5~mm thickness and 5.08~cm diameter is coated with aluminum wire grid on both surfaces with alignment to attain near uniform electric field inside the wafer (pointing along the y direction) and to retain optical transparency. Photolithography with lift-off process is used to deposit 100~nm thick aluminum grid wire on an area of 2.04~cm diameter and centered on both front and back sides of the wafer through back side alignment. Each aluminum wire is 4~$\mu$m thick and  separated by 40~$\mu$m. Wirebonding is used from the top and bottom electrode connections stretching from the center part coated with aluminum wire grid to the side of the wafer to connect to a PCB plane. The wafer is supported on the PCB through the use of three nylon washers which are equally separated and clamp the wafer from the sides. The washers hold the wafer through epoxy. The prototype ToF imaging system is shown in Figure~\ref{Measurement_Setup}.

\subsection{Mechanical Response}
\begin{figure}[H]
\begin{center}
\includegraphics[width=1\textwidth]{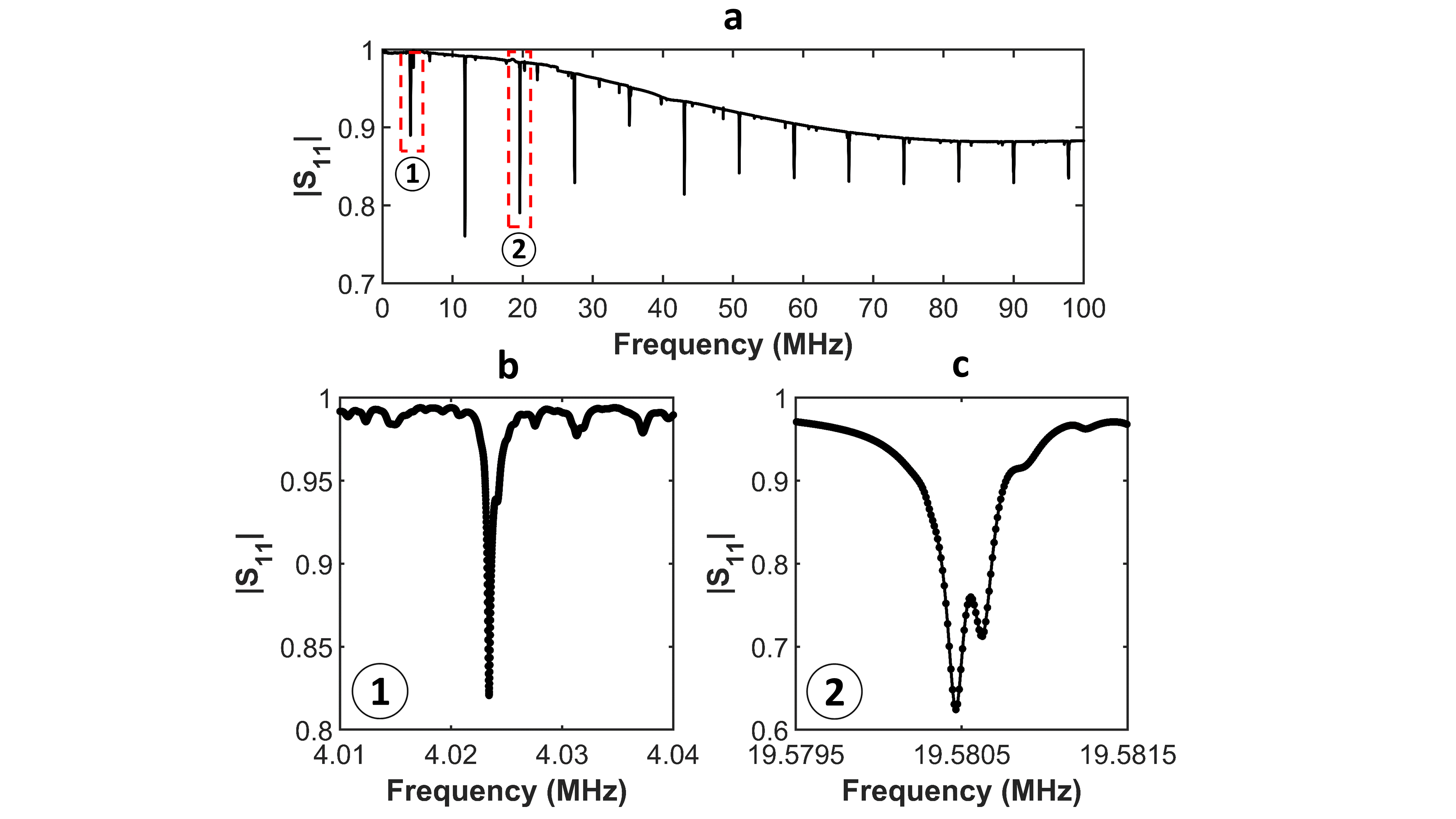}
\caption{(0.1-100)~MHz scan of $|S_{11}|$ for the fabricated lithium niobate wafer with respect to 50~$\Omega$. The fundamental mechanical resonance frequency appears at around 4.02~MHz and a higher order mode around 19.58~MHz. \textbf{a} $|S_{11}|$ with respect to 50~$\Omega$ of the wafer from 0.1~MHz to 100~MHz. \textbf{b} $|S_{11}|$ with respect to 50~$\Omega$ around the fundamental mechanical resonance frequency of the wafer, showing the resonance. \textbf{c} $|S_{11}|$ with respect to 50~$\Omega$ around the high order mode of the wafer appearing at 19.58~MHz, showing the resonance.}
\label{VNA_Scan}
\end{center}
\end{figure}

The mechanical response of the device is measured using a vector network analyzer (VNA). Figure~\ref{VNA_Scan} shows the mechanical frequency response for the device ($S_{11}$ parameter measured with respect to 50~$\Omega$). The fundamental mechanical resonance frequency shows up around 4.02~MHz and the other resonance modes are spaced by around 8~MHz, double the fundamental resonance frequency. The wafer supports modes up to 100~MHz, but the focus for the rest of this section will be on the fundamental mechanical resonance frequency at 4.02~MHz and the higher order mode at 19.58~MHz. We know from COMSOL simulations in Figure~\ref{1-100MHz_Complete} that these modes should have a net volume average for strain inside the wafer (corresponding to the COMSOL modes at 3.7~MHz and 20.5~MHz, respectively).  

\subsection{Optical Mixing}
\begin{figure}[H]
\begin{center}
\includegraphics[width=1\textwidth]{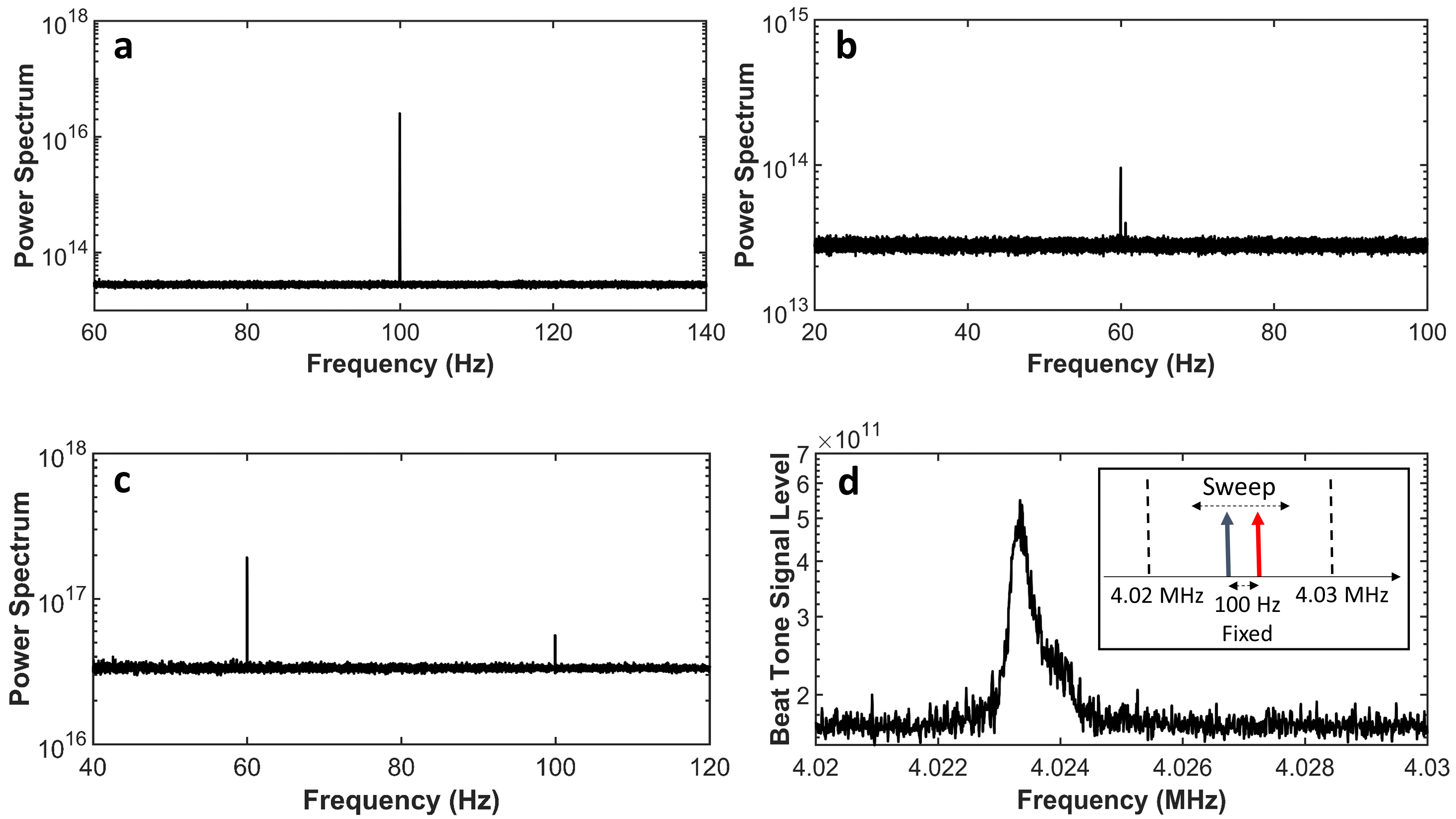}
\caption{\textbf{Optical mixing observed using the photoelastic modulator and captured with the Basler Ace 2040-90um CMOS image sensor}. An LED emitting light of wavelength 630~nm is used as the modulated light source. The wafer is driven with 20~V peak-to-peak and the camera has a frame rate of 400~Hz. 175,000 frames are captured and 576 image sensor pixels are spatially averaged to improve the SNR. \textbf{a} The wafer is driven at its fundamental mechanical resonance frequency at 4.0234~MHz, the LED amplitude modulation frequency is offset by 100~Hz from 4.0234~MHz and directed to the receiver (optical mixer placed in front of the camera). The beat signal at 100~Hz is visible after plotting the magnitude squared of the Fourier transform of the measured frames. \textbf{b} The wafer is driven at its higher order mode at 19.58097~MHz, the amplitude modulation for the LED is offset by 60~Hz from 19.58097~MHz and directed to the receiver (optical mixer placed in front of the camera). The beat signal at 60~Hz is visible after plotting the magnitude squared of the Fourier transform of the measured frames. \textbf{c} 4.023527~MHz and 19.58967~MHz are used simultaneously to drive the wafer, the LED amplitude modulation frequencies are 4.023587~MHz and 19.58977~MHz with the beat signals occurring at 60~Hz and 100~Hz, respectively. 100,000 image sensor pixels are spatially averaged to improve the SNR. \textbf{d} The frequency difference between the LED and the photoelastic modulator is fixed at 100~Hz and both frequencies are swept around the fundamental mechanical resonance frequency from 4.02~MHz to 4.03~MHz. The beat signal level appearing at 100~Hz is plotted, which shows that the photoelastic effect is dominant over the electro-optic effect since resonance behavior is observed.}
\label{Beat_Detection}
\end{center}
\end{figure}

To observe optical mixing on the CMOS image sensor and downconvert megahertz level amplitude modulation frequencies down to hertz range, we amplitude modulate a light-emitting diode (LED) emitting light of wavelength 630~nm at a frequency slightly offset from the mechanical resonance frequency of the wafer. The light passes through the optical mixer, which includes the aluminum deposited lithium niobate wafer. The system includes the amplitude modulated LED, polarizer, aluminum deposited lithium niobate wafer (photoelastic modulator) driven at one or more resonance frequencies, a 90$^\circ$ rotated lithium niobate wafer, and another polarizer. We observe optical mixing at 4.02~MHz when the wafer is driven at resonance and the LED is detuned in frequency by 100~Hz. We also observe mixing when the higher order mode is driven at around 19.58~MHz and the LED is detuned by 60~Hz. Multi-heterodyne detection is observed, in which two tones are driven simultaneously (4.02~MHz and 19.58~MHz) and the beat tones placed at 60~Hz and 100~Hz, respectively. The mixing terms are shown in Figure~\ref{Beat_Detection}. The figure also shows that the photoelastic effect is what causes the optical mixing, because when the frequency supplied to the photoelastic modulator is swept around the fundamental mechanical resonance frequency, the beat tone signal level (appearing at 100~Hz) changes and shows a resonance behavior.

\subsection{Analyzing Experimental Results}
The depth of intensity modulation is 0.1\% when 20~V peak-to-peak is applied to the wafer. The fundamental mechanical resonance mode at 4.02~MHz has a quality factor of roughly 11,000. The depth of modulation for the LED was 12\%. Using \eqref{Eq.15} and \eqref{Eq.20}, the expected depth of modulation can be calculated as 1.75\%.  

The discrepancy between the expected and measured depth of modulation could be due to the misalignment between the wafers (leading to constructive and destructive interferences as a result of static polarization). Another possible source could be the operation of the photoelastic modulator as an open-loop system. Since the fundamental mechanical mode has a high quality factor, to achieve high modulation depth the device needs to be operated at resonance, and even small frequency drifts in the fundamental mode should be tracked with a closed-loop system (e.g. phase-locked loop).

The observed optical mixing shows that the photoelastic modulator is a promising optical mixer. Depth of modulation can be improved through closed-loop driving to track any resonance drifts, aligning the optical components and re-fabricating the device to attain higher mechanical Q. Future work will focus on using the imaging system to form depth images in a scene. 

\section{Conclusion}
The working principle of a prototype phase-shift based ToF imaging system using an optical mixer, consisting of a photoelastic modulator sandwiched between polarizers, and placed in front of a standard CMOS image sensor is demonstrated. The photoelastic modulator is a Y-cut lithium niobate wafer, which has a thickness of 0.5~mm and a diameter of 5.08~cm. The photoelastic modulator is significantly more efficient than an electro-optic modulator for polarization modulation owing to the high mechanical Q and the strong piezoelectricity and photoelasticity of lithium niobate. The working principle of the system, including polarization modulation through the resonant photoelastic effect, converting polarization modulation to intensity modulation, and multi-frequency operation by simultaneously driving the photoelastic modulator at multiple of its mechanical resonance frequencies are demonstrated. We have demonstrated that with the addition of a cost-effective, compact optical mixer, a standard image sensor can function as a high resolution flash lidar system. 

\setcounter{secnumdepth}{0}
\section{Acknowledgments}
This work is supported in part by Stanford SystemX Alliance, Office of Naval Research, and NSF ECCS-1808100. Device fabrication was performed at the Stanford Nano Shared Facilities (SNSF) and the Stanford Nanofabrication Facility (SNF), supported by the National Science Foundation under award ECCS-1542152.

\bibliographystyle{ieeetr}
\bibliography{ieee_ref_Okan}

\end{document}